\documentclass[preprint2]{aastex}
\usepackage{graphicx}

\shorttitle{A search for hot massive extrasolar planets with NACO}
\shortauthors{Masciadri et al.}

\begin{document}


\title{A search for hot massive extrasolar planets around nearby young stars with the adaptive optics system NACO\footnote{Based on observations collected at the European Southern Observatory, Chile. Program $70.C-0777D$, $70.C-0777E$ and $71.C-0029A$}}

\author{E. Masciadri, R. Mundt, Th. Henning, C. Alvarez}
\affil{Max-Planck Institut f\"ur Astronomie,
K\"onigstuhl 17, D-69117 Heidelberg, Germany}
\email{e-mail contact:masciadr@mpia.de}
\author{D. Barrado y Navascu\'es}
\affil{Laboratorio de Astrof\'\i sica Espacial y F\'\i sica Fundamental, INTA, Apdo. Postal 50727, E-28080 Madrid, Spain }





\begin{abstract}
We report on a survey devoted for the search of exo-planets around young and 
nearby stars carried out with NACO at the VLT. The detection limit 
for $28$ among the best available 
targets vs. the angular separation from the star is presented. 
The non-detection of any planetary mass companion in our survey is used to 
derive, for the first time, the frequency of the upper limit 
of the projected separation planet-stars. 
In particular, we find that in $50$$\%$ of cases, no $5$M$_{J}$ planet (or more massive) has been detected at projected separations larger than $14$ AU and no $10$M$_{J}$ planet (or more massive) has been detected at projected separations larger than $8.5$ AU. In $100$$\%$ of cases, these values increase to $36$ AU and $65$ AU respectively. The excellent sensitivity reached by our study leads to a much lower upper limit of the projected planet-star separation compared with previous studies. 
For example, for the $\beta$ Pictoris group, ($\sim$$12$ Myr), we did not detect any $10$M$_{J}$ planet at distances larger than $15$ AU. A previous study carried out with $4$ m class telescopes put an upper limit for $10$M$_{J}$ planets at $\sim$$60$ AU. 
For our closest target (V$2306$ Oph - d = $4.3$ pc) it is shown that it would be possible to detect a $10$M$_{J}$ planet at a minimum projected separation from the star of $1$ AU and a $5$M$_{J}$ planet at a minimum projected separation of $3.7$ AU. 
Our results are discussed with respect to mechanisms explaining planet formation and migration and to forthcoming observational strategies and future planet finder observations from the ground.
\end{abstract}




\section{Introduction}
\label{intr}

At the present time more than $100$ exo-planets are known 
 which have been mostly discovered by 
the indirect method of the variation of the radial velocity induced 
by planets orbiting around the central star \footnote{Geneva Chronology Catalog 2004 - \newline \mbox{http://www.obspm.fr/encycl/catalog.html}}. 
It can be shown that young planets can be detected with adaptive optics (AO) 
imaging techniques at $8$-$10$ meter telescopes if they are sufficiently 
distant from their parent star. Since it is reasonable to assume
a coevality between the parent star and 
the planet, we can retrieve from atmospherical models 
(Burrows et al. 1997, Baraffe et al. 2003) that planets having a mass 
in the range $3$-$10$ M$_{J}$ and orbiting around young late-type stars 
($10$-$200$ Myr) have a typical brightness contrast 
with respect to the 
parent star of the order of $10^{2}$-$10^{6}$, corresponding to a magnitude difference of
$\Delta M$ of $5$-$15$ mag. This means that a planet can 
be detected at a few tens of AUs from parent stars having distances
$\leq$ $50$ pc. 
First results on the search for young and massive planets are reported by
Neuh\"auser et al. (1997) using 4 m class telescopes and  
 Macintosh et al. (2001) and Keisler et al. (2003) using 10 m class telescopes.
Recently (Chauvin et al. 2004), a companion of $\sim$ $5$$\pm$$2$M$_{J}$ was detected at around $50$ AU from the central star 2M1207, a very late type star (M8) in TWA Hydrae ($\sim$ $10$ Myr). This object is, at the present time, the most interesting giant planet candidate ever detected and further confirmation of the orbital motion is needed.  

From the point of view of the scientific motivations related to this 
kind of search
it is certainly an important challenge to image directly 
a planet around another star, since it can provide a direct evidence 
for the existence of such a planet. At the present time 
only a few exo-planets have been detected with the 
photometric technique called {\it 'transit'} (HD209458b - Charbonneau et al. 2000, 
OGLE-TR-113 (Konacki et al., 2004, Bouchy et al. 2004), OGLE-TR-56 (Torres et al., 2004), OGLE-TR-132 (Bouchy et al. 2004), OGLE-TR-111 (Pont et al., 2004) and TrRes-1 (Alonso et al., 2004)). Moreover, 
direct imaging is 
complementary to other detection methods like the radial velocity 
variation technique.  
The latter is sensitive to objects orbiting at close distances 
from the parent star (the radial velocity induced by the presence of 
a companion 
is indeed proportional
to the inverse of the square root of the orbital distance) while the 
AO imaging presented here would be sensitive to planets 
orbiting at much larger distances. 

According to statistical results retrieved by radial velocity studies (Marcy et al. 2003), 
most of the known exo-planets orbit
at distances smaller than $1$ AU and the semi-major axes of 
exo-planets found so far are 
not larger than $\sim$$6$ AU. Since the radial velocity technique is heavily biased 
towards short separations, the derived major-axis distribution is certainly 
not representative.
Not much is known at the present about potential planets that could 
exist at distances larger than about $6$ AU. 
We note that the formation of exo-planets is quite unlikely at distances 
of a few tens of AU. At these large distances the density of the planetesimals is low and the gravity from the central star is weak. This makes the process
of planet formation inefficient and slow (Thommes et al., 1999, 2002). However, different models predict the transfer 
(jump) or migration of planets to distances larger than $5$-$10$ AU from the central 
stars. Among these models, we recall the gravitational 
scattering model which claims that in multi-planetary systems, 
due to gravitational interactions between the planets, it can happen 
that the lowest mass objects are 
ejected on hyperbolic trajectories and some planets can move on stable 
orbits of a few tens of AU from the parent star. 
Such models were proposed by Weidenschilling $\&$ Marzari (1996) and 
Rasio $\&$ Ford (1996) and revised more recently by 
Papaloizou \& Terquem (2001) and Ford et al. (2001). 
An alternative mechanism that could induce an outward migration is the 
planet-disc interaction (Veras \& Armitage, 2004). It was proven that 
stable orbits of a few tens of AU can be attained by a few Jupiter 
mass planet under the effect of torques induced by the disc (i.e at 
early stages of the star formation) if the system has 
photoevaporation wind. Numerical simulations (Veras \& Armitage, 2004) 
showed that, under the assumption of a typical solar-mass photoevaporation 
wind of the order of $1$-$5$$\times$$10^{-9}$ M$_\odot$yr$^{-1}$, an outward 
migration of a planet with few Jupiter masses can take place with a frequency 
of the order of $25$ $\%$ (with respect to a sample of a few hundreds of simulations). 
It is, therefore, interesting to find out
if young and massive planets are present up to a few tens of AU from 
the central star.

For these reasons, we planned and performed an extended survey 
aiming to search for 
massive exo-planets orbiting around nearby and young stars 
using one of the best available instruments at the present 
time for high-contrast AO imaging observations: NACO at the Very Large Telescope (VLT) (see 
Section \ref{naco}).

The plan of the paper is the following. In Section \ref{targets} 
the criteria used for the selection of the targets are presented.
In Section \ref{naco} we describe the observational strategy and we 
provide technical information on the observations. In Section \ref{proc} 
we present our data processing, in Section \ref{planets} 
we discuss an analysis of planet-like features identified in deep images 
and in Section \ref{sensit} we give a detection limit estimate 
for the observed targets. In Section \ref{disc} we discuss our 
results and finally, in Section 
\ref{conc} the conclusions of our study are presented.

\section{Target selection}
\label{targets}

The properties of planets that we are searching for are: {\bf (a)} the contrast $\Delta M$ between the planet and the central star should be
$5$-$15$ mag {\bf (b)} and the planets 
should orbit at distances $a$ $\ge$ $1$ AU from the central star.


The planet intrinsic luminosity, retrieved by evolutionary 
models (Burrows et al. 1997, Baraffe et al. 2003), depends on the 
age of the system and 
the mass of the planet. The magnitude of a planet depends also on 
the wavelength used for 
the observations. Due to the fact that planets strongly fade with age we 
selected stars as young as possible.  
Following the evolutionary models, we define our 
'selection' ranges as follows: age $\sim$ $10$-$200$ Myr, mass of the planets 
$\sim$ $3$-$10$ M$_{J}$ and wavelengths in the $H$ and $K_{s}$ bands. 
Obviously the nearer the target is from the observer, 
the smaller is the distance from the central star that we 
can reach and probe. Therefore, we gave priority to nearby objects. We note 
that most ($\sim$ $70$ $\%$) of our targets have distances $\le$ $32$ pc and only $3$ objects are located between $50$ and $77$ pc. 
In order to optimize the contrast conditions, we gave 
priority to late-type stars (spectral type K-M). Only one target (HD $155555$) 
has a spectral type G5IV.
Table \ref{tab1} and \ref{tab2} summarizes the properties of the selected targets 
and the theoretical magnitude of planets that might orbit around them. The magnitude of the stars in the $H$ and $K_{s}$ bands are retrieved from the magnitude in V band and the spectral type (Simbad Astronomical Database\footnote{\mbox{http://simbad.u-strasbg.fr/}}).   
Table \ref{tab1} gives also the references from which the targets were selected.
A relatively large fraction of targets belong to the TW Hydrae local associations, $\beta$ Pictoris moving groups and Tucana/Horologium association (Zuckerman et al. 2001a, 2001b; Song et al. 2003, Barrado y Navascues et al. 1999, Barrado y Navascues 2001). The age of TW Hydrae is estimated to be $\sim$ $10$ Myr (Stauffer et al. 1995), the age of $\beta$ Pictoris is $\sim$ $12$ Myr (Barrado y Navascues et al. 1999) and the Tucana/Horologium age is $\sim$ $30$-$40$ Myr (Zuckerman \& Webb 2000). Several other targets, members of young stellar kinematic groups, were selected from Montes et al. (2001). 

We want to emphasize that the ages given in Table \ref{tab1} are often approximate values, particularly for those stars which are not members of associations or moving groups. For sufficiently young low-mass stars ($\le$ $30$-$50$ Myr) their location in theoretical Hertzsprung Russel diagrams can be used for an age estimate. For older stars ($\ge$ $50$ Myr) the isochrones are often indistinguishable from stars on the ZAMS (see for example Barrado y Navascues 1998) and therefore other age calibrators have to be employed like the stellar magnetic activity, the lithium abundance, the stellar rotation rate and the X-ray flux. In addition to these photometric signatures, the galactic space velocity is often employed for an age estimate and for defining young moving groups. Young stars are indeed characterized by low dispersion of the galactic space-velocity components (U, V, W). A few criteria exist in the literature 
defining precise ranges for these components
(Eggen 1996, Jeffries 1995). 
These same criteria are employed by Montes et al. (2001). 

We avoided the selection of binary systems with 
separations smaller than $\sim$ $2$$\arcsec$ because 
the probability 
that a planet can find a stable orbit around one of the binary component is
rather small. 
The binary systems included in our survey are: TWA $5$ ($2$$\arcsec$), 
TWA $8$ ($13$$\arcsec$), TWA $9$ ($6$$\arcsec$), GJ $799$ ($2\farcs{8}$) and 
BD $-17^{\circ}$ $6128$ ($2\farcs{2}$).

\section{NACO Observations}
\label{naco}

The observations were performed with the AO system NACO, consisting of the infrared camera CONICA (Lenzen et al.\ 1998) assisted by the adaptive optics facilities NAOS (Rousset et al.\ 2000) placed at the focus of the unit ``Yepun'' of the ESO/VLT $8$ m telescope.
We observed $30$ targets during three observing runs: 16-19 February 2003, 20-23 July 2003 and 7-9 January 2004. Two targets (GJ~$182$ and GJ~$179$) were observed in service mode. Table \ref{tab3} summarizes the observational instrumental parameters.

Due to the rather bright central star, our observational strategy was the following. 
First, we observed each target in a narrow-band filter (Table \ref{tab3}) for a few minutes
in order to obtain a non-saturated point spread function (PSF). We obtained a Strehl ratio (SR) in the range $0.2$-$0.55$ over the all runs. 
Second, we obtained deep and saturated ($\sim$ $20$-$30$ min) 
$K_{s}$ or $H$ broad-band observations of each target (Table \ref{tab3}) in order to increase the detectability 
on the PSF wings. 
We used NACO at a pixel scale of $13.25$ mas yielding a FOV of $14\arcsec\times14\arcsec$. 
The optical wavefront sensor was locked on the target star yielding a typical resolution of $\sim$$50$ mas in $K_s$ band. 

For the deep observations, we selected the exposure time (DIT in Table \ref{tab3}) of a single frame
 to obtain a number of counts (for the central star) larger 
than $\sim$ $10-30$ times the 
saturation limit. The deep images are saturated over a central diameter 
having a size of $0\farcs{1}$-$0\farcs{2}$. For each target, a data set of dithered frames was obtained 
in order to correct for sky emission and the pixel to pixel variations in sensitivity.

In a few cases (e.g. targets characterized by a smaller brightness) we used a broad-band filter with a neutral density filter (ND Short - see Table \ref{tab3}) instead of a narrow-band filter to obtain a non-saturated PSF. 
For HD $17925$, due to its high brightness, we used a different observational strategy based on the absorption 
properties of a planet atmosphere. This target was observed with the intermediate-band filters (IB $2.09$ and IB $2.24$) centered out and in the methane (CH$_{4}$) absorption band ($\sim$ $2.24$ $\mu$m). We expect that planet features would appear in the continuum (IB $2.09$) but not in the absorption band (IB $2.24$). 

Due to the higher Strehl ratio (SR) reachable in the $K_s$ band compared to the $H$ band,
we gave priority to the former band in order to optimize the probability
to detect a low-mass companion. We chose the $H$ band only for the 
oldest stars ($100$-$200$ Myr), i.e. for planets having a 
theoretical colour 
($H$-$K$) $\lesssim$ $-$ $1.3$ mag. We note that, according to 
the COND models (Baraffe et al. 2003) (see Fig.\ref{fig1}) and for planet masses of $3$-$10$ M$_{J}$, the $H$ band is comparable to the 
K band for an age minor than $100$ Myr. For targets with an age of 
$100$-$200$ Myr the gain of the H band with respect to the K band 
is significant.


\section{Analysis}
\label{anal}

\subsection{Data processing}
\label{proc}

The procedure used to reduce our data set includes IRAF\footnote{IRAF is distributed by the National Optical Astronomy Observatories, http://iraf.noao.edu/} and ECLIPSE\footnote{ECLIPSE is a data reduction package developed by ESO, N. Devillard, "The eclipse software", The messenger No 87 - March 1997} packages. Each frame was cleaned by a bad-pixel mask and divided by the flat field. In addition, the
sky background was eliminated in each frame. The final deep image (shift and add of all the frames) was obtained using a cross-correlation routine in ECLIPSE. We underline that the shift and add had to be done using a cross-correlation method since our PSFs are saturated in the central zone. Alternative methods, based on the gaussian fit to the peak of a PSF fail in such a case. 

In order to save observational time, we did not observe any 
reference standard star 
but we filtered the final deep images from the 
low spatial frequencies using two 
different methods.
In this way the large-scale features of 
the saturated PSF were removed.

In the first method, we convolved the deep image of 
the scientific target
with a two-dimensional Gaussian function having a width roughly 
equivalent to the
FWHM of the PSF of the star. Then we subtracted 
the result from the original deep image. 
Only features identified outside the central saturated region ($\sim$ $0\farcs{1}$-$0\farcs{2}$) are reliable in all cases.
We note that, if the FWHM of the Gaussian function is much smaller than the 
FWHM of the PSF of the star, the filtering is not efficient. This means that large-scale features of 
the saturated PSF are not sufficiently eliminated and the potential planet-like features close to the star 
are not visible. If, on the contrary, the 
FWHM of the Gaussian function is too large, the filtering eliminates also 
the potential planet-like features close to the star i.e. the structures having small 
spatial scale. 
 
In the second method, we selected, for each point of the deep image, an annulus. Calculating the average of these frames we obtain the estimate of the background of the deep image inside an annulus around the 
centre of our target. We finally subtracted the background from the original image. 
The annulus is characterized by an inner radius of half of the 
PSF $FWHM$ and an outer 
radius $\sim$$3$ larger than the internal radius. The central part of the PSF is excluded in the calculation of the background because it introduces an offset in the average estimation. Reducing the size of the inner radius, we increase the filtering of the low spatial frequencies. Increasing the outer radius the estimation of the background is affected by features belonging to regions far away from the central part of the PSF. We define an outer radius equal to $3$ times the inner radius in order to estimate a background on a limited region around the central part of the PSF. We note that the final detection limit only weakly depends on the values of the inner and outer radius. Repeating the filtering with several combinations of inner and outer radii the final detection limit shows differences within $0.2$ mag. 

The filtering was applied in a region of 
$\sim$ $6$$\arcsec$$\times$$6$$\arcsec$ centred on each target.
Although the two methods have differences, they are based on the same principle aiming to eliminate the large-scale features of 
the saturated PSFs. In Section \ref{sensit} the similarities/differences of the two methods are discussed with respect to the results that we obtained.

\subsection{Point-like features}
\label{planets}

After data reduction and filtering of all the observations as described in Sec.\ref{proc}, a visual inspection of all the deep images was performed and a few point-like structures were identified in some 
of the images: V$343$ Nor ($K_s$ band), SAO $252852$ ($H$ band), GJ $179$ ($H$ band), BD $+2^{\circ}$ $1729$ ($H$ band), BD $+1^{\circ}$ $2447$ ($H$ band), GJ $183$ ($H$ band) and HD $221503$ ($H$ band). All the other targets did not reveal any point-like sources around them. We discuss here the nature of these objects trying to distinguish between artifacts (ghost images) and faint point sources at low galactic latitude.\newline

We have verified that all the targets observed in H band show three point-like features at the same position at about $1$$\arcsec$ from the star. 
Fig.\ref{bd_filter} shows deep images of two of these targets: BD $+2^{\circ}$ $1729$ and BD $+1^{\circ}$ $2447$. Three point-like features are visible (indicated by arrows and letters A, B and C) at the same relative angular separation from the star. These features are probably 'ghost' images produced by some imperfection of the broad-band filter or static aberration residuals that depend on the wavelength. This could justify the fact that they are visible only in $H$ band.\newline   

In only two deep images (V$343$ Nor and SAO $252852$ - Fig.\ref{sao_cand})  we observed clear point-like sources different from those created by the $H$ filter that we have just described. Fig.\ref{sao_cand}a shows the deep image of SAO $252852$ (galactic latitude = $-4.6^{\circ}$). An object at around $25$-$30$ AU from the star is visible. The estimated magnitude of this object is $\sim$$17$ mag in $H$ band. Due to the low galactic latitude of SAO $252852$, the probability that this feature is a background star is not negligible.  
Fig.\ref{sao_cand}b shows the deep image of V343 Nor (galactic latitude = $-1.9^{\circ}$). 
Several sources were detected around this star. The estimated magnitude of all the sources is in the range $15$-$20$ mag. We note that a $5$M$_{J}$ planet orbiting around this $12$ Myr old star would have $\sim$ $16$ mag in $K$ band. Besides this, we observe that the galactic latitude of V$343$ Nor is relatively low ($-1.9$) and several among these point-like sources appear a few arcsec away from the central star. Thus the probability that these sources are background stars is quite high. 
We can not totally exclude that the faint sources around SAO $252852$ and V$343$ Nor are gravitationally bound massive planets. For this reason, a second epoch observation is necessary in order to clarify the nature of the detected point-like sources. Knowing that the proper motions (Hipparcos catalog, Perryman et al. 1997) of SAO $252852$ are ($\mu_{\alpha}$$=$ -202  mas/yr and $\mu_{\delta}$$=$ -270 mas/yr) and those of V$343$ Nor are ($\mu_{\alpha}$$=$ -$52.87$ $\pm$ $1.16$  mas/yr and $\mu_{\delta}$$=$ -$105.99$ $\pm$ $0.98$  mas/yr) we deduce that after one year a re-observation could throw light on the nature of these objects. A forthcoming run is planned to definitely verify if these objects are indeed background stars. 
We remind the reader that an estimation of the probability that a point source having a faint magnitude comparable to ours in a field of view of $2$$\arcsec$ is a background star was done for a galactic latitude of $\sim$ +$15^{\circ}$ (Brandner et al. 2000). A probability of $70$$\%$ was estimated in that case. Our point sources are placed at a lower galactic latitudes ($-1.9^{\circ}$ and $-4.6^{\circ}$), and the point-sources are placed at around $2$$\arcsec$ from the central star (in the SAO $252852$ case, Fig.\ref{sao_cand}) and at distances larger than $2$$\arcsec$ (in the V$343$ Nor case, Fig.\ref{sao_cand}). We conclude, that the probability that such sources are background stars is larger than $70$$\%$.

We conclude, thus, that no very promising massive-planet candidate was identified in our survey. 

\subsection{Estimate of detection limits}
\label{sensit}
 
To estimate quantitatively the detection limit of our deep images as a function of the angular separation from the central star, we used the following procedure.
Along a radial direction we calculated the standard deviation of the intensity
over a box of $d$$\times$$d$ pixels (with typical values of $d$ of $4$-$6$ pixels which corresponds to the FWHM of the non-saturated PSFs) and one pixel step.
We then averaged over all radial directions ($360$$^{\circ}$) and 
finally we calculated the contrast $\Delta M$ at $5$$\sigma$ with respect 
to the peak of the non-saturated PSF of the star. Since the central star is saturated in the broad-band images we had to use the fluxes measured in the non-saturated narrow-band filters (see Table \ref{tab3}) to calculate what peak flux our saturated broad-band images would have if they were not saturated. The non-saturated PSF had to be multiplied by a constant factor ($A$) to take into account the different transmission and width of the filters. If $F_2$ is the filter used to obtain the saturated PSF and $F_1$ the one used to obtain a non-saturated PSF, $A$ is given by:

\begin{equation}
A=\frac{FWHM_{F2}}{FWHM_{F1}}\cdot \frac{Tr_{F2}}{Tr_{F1}}\cdot \frac{%
DIT_{F2}}{DIT_{F1}}
\end{equation}
where $Tr$ is the filter transmission coefficient, $FWHM$ is the full width half maximum of the filters and DIT is the integration time of each single frame (Table \ref{tab3}).

In order to illustrate that our detection limit is reasonable, we scaled the non-saturated PSF of one of our targets (V343 Nor) to a flux value corresponding to a $5$M$_{J}$ planet (representative of the typical planets we are interested in) and we placed this 'artificial planet' in the deep image at different angular separations from the star. Fig.\ref{mascia}~(a)-(b) shows the filtered deep images of the same target processed with the two filtering methods described above (Section \ref{proc}). This ``artificial planet'' is placed at $1\arcsec$ and $0\farcs{7}$ from the central star. Fig.\ref{mascia} (a)-(b) shows that a separation of $0\farcs{7}$ is roughly the minimum angular separation at which a $5$M$_{J}$ planet can be detected. 

Figures \ref{det_lim1}-\ref{det_lim4} show the detection limits as a function of angular separation for $28$ objects of our survey for which deep images have been obtained. We note that, in one case (GJ $207.1$), a technical problem did not allow to observe the non-saturated PSF and this prevented us to calculate the detection limit. In the case of the binary GJ $799$, we could calculate only the detection limit for GJ $799$ A. The separation of $\sim$ $2\farcs{8}$ between the two components was slightly too large to include both stars in the same frame in all the dithered positions. 

For all targets, the detection limit is reliable starting for angular separations $\ge$ $0\farcs{1}$. For angular separations $\le$ $0\farcs{1}$ the PSF of the deep images is usually saturated. 
Fig.\ref{det_lim4} shows the detection limit of the two targets (HD $17925$ and BD -17$^{\circ}$ $6128$) that we observed through the more narrow ``methane band filters'' (see Section \ref{naco}). Since the optical quality of these intermediate-band filters is worse compared to the broad- and narrow-band filters, the final deep images appear dominated by structures characterized by high spatial frequencies. For this reason, the filtering is less efficient for these two deep images (the dashed, dotted and full thin lines are quite similar - Fig.\ref{det_lim4}).

From Fig.\ref{det_lim1}, \ref{det_lim2}, \ref{det_lim3} we can see that:
\begin{itemize}
\item The detection limit of the filtered deep images is very similar for the 
two different filtering methods. This indicates that the achieved gain from the applied 
'filtering' is practically method independent. 
\item The filtering of large spatial structures (low spatial frequencies) has a variable efficiency. The angular separation range over which the filtering shows
its maximum efficiency is $0\farcs{5}$-$1\farcs{5}$ from the central star. In this region the gain in contrast $\Delta M$ with respect to the non-filtered image ranges from $0.5$ to $2$ mag.
\item The detection limit is not the same for all the stars. This is not surprising because it depends on several parameters: the age, the magnitude of the central star, the wavelength used for the observations, the distance of the star from the observer and also on the exposure time. In one case (the brown dwarf TWA $5$B), the detection limit is rather low because we did not saturate the PSF since the central star is relatively faint.
\end{itemize}

To calculate the typical performances of NACO at subarcsecond distances, we averaged the contrast $\Delta M$ obtained at $0\farcs{5}$ and $1\arcsec$ for all the observed targets having a total integration time within the range $20$-$30$ minutes.
We find, in the $K_s$ band, a $\Delta M$ $=$ $9$ mag at $0\farcs{5}$ and a $\Delta M$ $=$ $11.5$ mag at $1\arcsec$. We find, in $H$ band, a $\Delta M$ $=$ $9.2$ mag at $0\farcs{5}$ and a $\Delta M$ $=$ $11.7$ mag at $1\arcsec$. We recall that $0\farcs{5}$ corresponds, for our sample, to projected separations of $2$-$35$ AU, $1\arcsec$ to projected separations of $4$-$70$ AU. 

Luhman \& Jayawardhana (2002), observed $25$ targets with NICMOS (Keck II) and found a $\Delta M$ $\sim$ $10$-$10.5$ mag at $1\arcsec$ in the $H$ band with an integration time of $\sim$ $18$ minutes. This illustrates nicely that 
NACO is a well performing instrument for high-contrast imaging observations.


\section{Discussion}
\label{disc}

\subsection{Upper limit of the projected separation between the star and a potential planet}
\label{dist}

The detection limit estimates also permit us to calculate 
an upper limit of the projected separation between a star and a potential planet, as follows.
We consider a typical $5$M$_{J}$ planet,
we calculate the theoretical value (retrieved by the models) 
of the star/planet contrast, we then place this value on the ordinates of the detection limit curve (see for example Fig.\ref{det_lim1}) and finally we retrieve (on the abscissae) the distance beyond which we can affirm that such a planet 
should not exist (because it was not detected). Table \ref{tab4} shows, for each target, such upper limits for the star/exo-planet distance. Table \ref{tab5} shows the frequency at which we would be able to detect a $5$M$_{J}$ planet at different distances from the parent star. 
By analysing these results from a statistical point of view, we can 
calculate (Fig.\ref{prob}) the cumulative distribution 
of the upper limit of the distance star/exo-planet for our sample of targets. The thin line represents the cumulative distribution obtained for $5$M$_{J}$ planets. We find that (Fig.\ref{prob}), with respect to our sample of targets, in $50 \%$ of cases (median value), there are not planets at distances larger than $14$ AU and, in $100 \%$ of cases, there are no planets at distances larger than $65$ AU. We underline that we are considering a $5$M$_{J}$ planet just as a reference, but the same calculation could be done for more and/or less massive planets in the range $3$-$10$ M$_{J}$. In the case of more massive planets, the cumulative distribution should have a median value ($50 \%$) associated with an angular separation smaller than $14$ AU. In the case of less massive planets, the cumulative distribution should have a median value (50 $\%$) associated with an angular separation larger than $14$ AU. To better appreciate the sensitivity with respect to the mass of the planets we also calculated the cumulative distribution for $10$M$_{J}$ planets (Fig.\ref{prob} - bold line). We find that, in $50 \%$ of the cases (median value) there are no planets at distances larger than $8.5$ AU and in $100 \%$ of the cases there are not planets at distances larger than $36$ AU. Table \ref{tab4} and Table \ref{tab5} show the upper limit of the projected star/planet separation and statistical results also for $10$M$_{J}$ planets. 
The numbers given in these tables illustrate that it is relatively unlikely that a massive planet exists beyond a certain distance from the star. For example the fact that no $5$M$_{J}$ planet was found for $18$ of our targets beyond $20$ AU (see Table\ref{tab4}) means that the likelood for the existance of such planets beyond this distance is $\sim$ $(1/18)$ $\%$ i.e. $\sim$ $5.6$$\%$. This last estimation is useful for the reader to have an idea of how much the size of the sample affects the statistical results. We recall that we are using the term {\it probability} in a less conventional way. A precise (statistically speaking) definition of {\it probability} should need some positive detection that we obviously do not have.

Which are the implications of these results with respect to models such as those cited in Section \ref{intr} (gravitational scattering model and outward migration), explaining the transfer of planets to large distances (some tens of AU) from the central star ? Our results indicate that the frequency of planet 'jumps' or outward migration is quite low. More precisely, it seems difficult for $10$M$_{J}$ planets to find a stable orbit at distances larger than $16$ AU, and that $5$M$_{J}$ planets at distances larger than $26$ AU, at least in the range of early ages that we considered. 

We note that the statistical analysis presented in Fig.\ref{prob} related to the $10$M$_{J}$ planets is particularly interesting because, due to a steeper cumulative distribution concentrated at small projected separations from the central star, it permits us to better constrain the probability of location of planets. We also recall that such massive planets are rarely detected at distances smaller than $6$ AU using radial velocity measurements. As described in the introduction, the radial velocity technique is sensitive only to planets orbiting at closer distances ($\lesssim$ $6$ AU) from the central star. It is interesting to analyse what the direct imaging technique reveals about $10$M$_{J}$ planets at distances larger than $6$ AU. Looking at our results, we can state that, for most of the targets ($\sim$ $90 \%$ of cases), there are no $10$M$_{J}$ planets at distances larger than $\sim$ $16$ AU. We find that the probability to find planets equal or more massive than $10$M$_{J}$ at large distances is low as well as the probability to find them at close distances ($\le$ $6$ AU) as appears from radial velocity surveys (Marcy et al. 2003). Massive planets are expected to form beyond the so called {\it 'snow line'} distance ($\sim$ $5$ AU for the solar-like stars\footnote{We underline that the {\it 'snow line'} is equal to $5$ AU for typical solar-like star. This distance is calculated for simple thermal equilibrium and is estimated equal to $1$-$2$ AU for late type stars (Black 1980)}) (Lissauer, 1993) and from this distance they might migrate inward. 
Our study shows that, at least for late spectral type (K and M) stars, if such massive ($\sim$ $10$M$_{J}$) planets form beyond the snow line, we can envisage the following possibilities: {\bf (1)} massive planets form and can be found in the narrow range of distances {\it 'snow line'} - $16$ AU from the central star, {\bf (2)} at distances larger than $16$ AU only planets having masses smaller than $10$M$_{J}$ can form and can be found, {\bf (3)} if any $10$M$_{J}$ planet would have formed beyond the {\it 'snow line'}, the formation and inward migration necessarily happen on a time scale $\le$ $10$Myr. To test the hypotheses (1) and (2) one should improve the contrast at small angular separations. To test the hypothesis (3) one should select a sample of stars younger than ours. Unfortunately, stars having this property are placed preferably at distances $\ge$ $100$ pc from the observer. Even future ground-based planet finders will be able to detect planets only at larger separations (a few tens of AU) from the central star if the targets are placed at such distances ($\ge$ $100$ pc). For this reason, it would be probably more efficient and appropriate to observe them from space. In this way, one could by-pass the limitations imposed by the atmosphere on the attainable angular resolution and to probe the same projected separations (as those analysed in this paper) from these distant stars. The resolution reachable in space by the Hubble Space Telescope (HST) is not good enough to probe such small physical distances. In a previous study (Brandner et al. 2000), a set of pre-main sequence T Tauri stars in the Chamaleon T ($1$-$5$ Myr at $150$ pc) and Scorpius-Centaurus OB ($5$-$15$ Myr at $150$ pc) associations were observed. Results showed that at $30$ AU it would be possible to detect only more massive planets ($\sim$ $20$M$_{J}$) than those we would detect with NACO (this paper). We should thus wait for a new generation of space telescope to probe such close distances at so early ages. 

Other interesting issues can be retrieved by our analysis concerning the core-accretion model. Our results provide interesting information concerning mechanisms to explain planet formation. We are considering here the 
core-accretion model originally proposed by Pollack (1996) since it provides an explanation 
for the formation of all gaseous as well as solid planets\footnote{We precise that the core-accretion model was proposed for solar-like star and our sample is done by young stars havinging later spectral type than solar-like stars.}. It claims that a growing solid core, after reaching a critical mass, accretes a massive atmosphere. According to our results, this model could 
work, assuming that massive planets 
($\ge$ $10$M$_{J}$) form in a narrow $5$-$15$ AU distance 
from the central star in a time 
scale of $\sim$$10^{6}$ yr (typical time scale of disk lifetime). 
One of the critical aspects of the core-accretion model has always been that it can not explain the presence of planets observed with the radial velocity technique at close distances from the central star. The time necessary for the formation plus the migration should be longer than the typical disk lifetime ($\sim$$10^{6}$ yr, Haisch et al. 2001). The modified version of the Pollack model, recently proposed by Alibert et al. (2004), 
seems to solve this problem. Using a detailed description of the composition (gas plus solid) of the disk this model reduces the time necessary for a planet to form and migrate by a factor of $10$. With such a modification, the process of formation and migration might need a time shorter than the typical disk lifetime. Following the predictions of Alibert's model, it would be possible that we did not detect massive planets simply because they formed at ages ($\le$ $10$Myr) earlier than those of our selected sample. This means that our results are consistent with the core-accretion model in the Alibert formulation. This model seems thus promising due to the fact that it matches with both observational issues coming from radial velocity measurements and also from direct imaging observations.

We finally mention that our results indicate that a sort of {\it 'exo-planet desert'} for massive ($\sim$ $10$M$_{J}$) planets is observed at distances larger than $15$ AU from the stars of our survey. This desert seems to extend to quite large distances from the central star. A recent survey (McCarthy $\&$ Zuckerman, 2004) carried out with the near-infrared camera NIRC and a coronograph at Keck did not detect any $5$-$12$ M$_{J}$ exo-planet in the range $75$-$300$ AU around $\sim$ $42$ stars having a mean age of $300$ Myr. 


 
\subsection{Level of sensitivity: comparison with previous surveys}
\label{sensitivity}

We would like to mention the relevant improvement in terms of sensitivity 
 reached by our study (using NACO at the VLT) with respect to previous 
published results. 
A recent analysis of massive planet detection (Neuh\"auser et al. 2003), carried out with the MPE Speckle Camera at the ESO's $3.5$ m NTT telescope, put an upper limit of $60$ AU on the distance between the star and $10$M$_{J}$ exo-planets in the case of targets belonging to $\beta$ Pictoris ($\sim$$12$ Myr). Results from our study show that we can exclude any $10$M$_{J}$ planets at distances larger than $15$ AU (a physical distance $4$ times smaller) for targets belonging to the same moving group. 
Our results show a better sensititity also with respect to previous observations made with HST (Neuh\"auser et al. 2002). Estimates provided by this study give an upper limit distance between the star and $5$M$_{J}$ planets of $\sim$ $50$ AU. Stars observed by Neuh\"auser et al. (2002) belong to a $\sim$$2$Myr old systems (well younger than our sample - $10$-$200$ Myr). Models predict a mean difference in H and K band of the order of at least $3$ mag between the age of stars in Neuh\"auser et al. study and ours. We thus deduce that the same instrument (HST) would have an upper limit for the star/planet distance quite a bit larger than $50$ AU for $5$M$_{J}$ planets in the $10$-$200$ Myr age range.

\subsection{Forthcoming observational strategy}
\label{sdi}

How can we improve our results to better constrain planet-formation models ?
It is evident that an increase of the achievable contrast at smaller angular separations shifts the median value of the cumulative distributions to smaller angular separations (Fig.\ref{prob}). This means that the upper limit of the projected star/exo-planet separation decreases. However, we note that, using the observational strategy employed in our survey, we can hardly improve the sensitivity at subarcsecond distances using a $8$ meter class telescope as illustrated in Fig.\ref{sdi_pot}. This figure shows the detection limit obtained for one of our targets for an integration time of $5$, $10$ and $22$ minutes. The deep image was filtered from low spatial frequencies before we calculated the detection limit. We observe that, beyond $1$$\arcsec$, the reachable contrast increases with increasing the integration time ($\sim$ $1$ mag in $15$ min). At distances smaller than $1$$\arcsec$, on the contrary, a longer integration time does not increase the contrast because we are strongly limited by the speckle noise.\newline 

The differential imaging technique proposed by Racine et al. (1999) is a rather efficient method to reduce the speckle noise at small angular separations from the central star. Images taken simultaneously at slightly different wavelengths are subtracted from each other. Since the speckle noise of both images is nearly the same\footnote{We precise that speckles of images taken at slightly different wavelengths match if frames are rescaled opportunely by a $\lambda_2$/$\lambda_1$ factor.}, the speckle noise is eliminated in the resulting image providing one can sufficiently limit static aberrations. The differential technique finds a useful application in detecting cool methane-rich faint objects. Rosenthal et al. (1996) first proposed 
observations of the same star done simultaneously
in two different wavelengths which are located {\it ``on''} and {\it ``off''}
the methane (CH$_{4}$) absorption band. If one subtracts the image taken in the absorption band 
from the one taken outside the absorption band,
 planet/brown dwarf features appear in the continuum. 
The absorption at $1.62$ $\mu$m, in the CH$_{4}$ band, is a clear signature 
of any object having temperature T $\leq$ $1300$ K (giant exo-planets and T-type
brown dwarfs). 

We can thus increase the sensitivity (contrast) at subarcsecond distances using the recently implemented NACO/SDI (Simultaneous Differential Imager) instrument. 
This instrument employs 
the differential technique in $H$ band. Preliminary results (Close et al. 2004, Lenzen et al. 2004) indicate a $\Delta M$ $\sim$ $11$ mag at $0\farcs{5}$ from the central star. Considering our averaged estimate of $\Delta M$ $=$ $9.2$ mag in the $H$ band (see Section \ref{sensit}), we should gain $\sim$ $2$ mag in contrast using NACO/SDI at $0\farcs{5}$ (see Fig.\ref{sdi_pot}). 

A further and fundamental improvement in contrast should be attained by 
future ground-based ``planet finders'', i.e. instruments supported by AO systems providing a better Strehl Ratio. 
The contrast reachable by such an instrument at $0\farcs{5}$ 
should be $\Delta M$ $\sim$ $17.5$ mag in $J$-$H$ bands 
(see Fig.\ref{sdi_pot}). We refer in particular to the top level requirements 
of CHEOPS, a second generation instrument for the VLT conceived for planet detection (Gratton et al. 2004). We think that the principal goals of these instruments (NACO/SDI and ground-based planet finders\footnote{We underline that, in the context of 
this paper, we are talking about the category of 
planets in which the intrinsic luminosity is quantitatively 
dominant over the reflected one.
Planet finders will also be able to detect colder and older 
exo-planets for which the reflected light is dominant with respect to the intrinsic one. This other category of planets orbit in 
general at closer distances 
from the central star.}) in the 
context of the detection of massive and young exo-planets and the 
understanding of mechanisms responsible for exo-planet formation 
are the following: {\bf(A)} They will permit us to provide answers to the 
hypothesis (1) and (2) (see Section \ref{dist}). 
  {\bf(B)} They will be rather appropriate to do studies similar 
to the one proposed in this paper but applied to a sample of targets characterized by an earlier spectral type (i.e. solar-like type) in the same range of ages. This is due to the better contrast reachable at subarcsecond distances (see Fig.\ref{sdi_pot}). 
{\bf(C)} They will be appropriate instruments to detect even fainter 
exo-planets (i.e. in the $1$-$5$ M$_{J}$ range) at subarcsecond angular separations from the central star.

\section{Conclusion}
\label{conc}

In this paper we have presented results of a survey carried out with NACO at the VLT aiming to detect massive exo-planets ($3$-$10$ M$_{J}$) orbiting around $30$ young ($10$-$200$ Myr) and nearby ($d$ $\le$ $77$ pc with $70$ $\%$ of them in the first $32$ pc) stars. 

No promising candidate was identified. All faint sources found in the vicinity 
of the central star are presumably background stars. A detailed estimate of the detection limit in $28$ of the observed targets was presented. This permitted us to calculate the typical contrast ($\Delta M$) between a planet and a central star at $0\farcs{5}$ and $1$$\arcsec$ provided by NACO for a typical integration time of $20$-$30$ min (see Section \ref{sensit}), to present a statistical analysis of the upper limit for the star/planet separation and also to discuss our results in the framework of the mechanisms that are supposed to explain the formation of exo-planets. 

Our most relevant result is that, in our sample of targets, in $100$$\%$ of cases, no $10$M$_{J}$ (and more massive) exo-planets was detected at distances larger than $36$ AU and no $5$M$_{J}$ (and more massive) exo-planets was detected at distances larger than $65$ AU. If one considers the median values of the cumulative distributions we have that, in $50$$\%$ of cases, no $10$M$_{J}$ (and more massive) exo-planets should exist at distances larger than $8.5$ AU and no $5$M$_{J}$ (and more massive) exo-planets should exist at distances larger than $14$ AU. We emphasize that, these statistical estimates were done for the projected star/exo-planet distance. We are not considering here the possibility that planets might not be visible in part of their orbit. Our study indicates that massive planets (mass $\ge$ $5$M$_{J}$) are rare at distances larger than $6$ AU as well as at distances smaller than $6$ AU as indicated by radial velocity estimates (see Section \ref{dist}).\newline
We underline that for the category of massive, warm and self-luminous planets that we looked for in our survey, it is not possible to calculate a reliable probability distribution for detecting planets. The reason is that, to do this, one should need to know the planet frequency distribution {\it 'a posteriori'}. This is obviously impossible to do if no-planets have already been detected. For this category of planets, one can only define some merit functions (depending on several parameters such as the age, the distance from the observer, the magnitude of the central star, the mass of the planet) to define some criteria to select targets for which the detection should be particularly favourable. For this reason, we think that it is fundamental to report results (even of non-detections) for the category of exo-planets that we are searching for and to try to get as homogeneous analysis as possible of the results in order to improve the statistics of the upper limit for star/planet distance.

\section{Appendix}
\label{app1}

{\bf GJ 803 (Au Mic):}
                                                   
Recent estimates of emission at optical wavelengths (Liu et al. 2004, Kalas et al. 2004) of a circumstellar disk surrounding GJ 803 (AU Mic) were published. The disk is detected between $\sim$ $50$ and $210$ AU corresponding to around $5$$\arcsec$ - $21$$\arcsec$ (d $\sim$ $10$ pc). The inner region was occulted by the presence of a coronograph so no information is provided on the disk structure at distances smaller than $5$$\arcsec$. The field of view of our observations is much more smaller ($\sim$ $2$$\arcsec$ in radius around the central star) than that used for the disk detection. Within $\sim$ $50$ AU, the typical time scale of dust is shorter than the stellar age so it was suggested (Kalas et al. 2004) that grains observed here must be continually replenished by the collisional erosion of much larger objects (like comets and asteroids). In the inner region of the disk, the presence of planetisimals might originate planets by accretion. Kalas et al. (2004) suggest that in the $2$ - $30$ AU region, the presence of gas giant planets should be possible. Our results indicate that no planets equal or more massive than $5$M$_{J}$ are detected at distances larger than $4$ AU and no planets equal or more massive than $10$M$_{J}$ are detected at distances larger than $2$ AU. Besides this, we can not exclude that some planets with mass smaller than $5$M$_{J}$ exist at such close angular separations. We have to wait for observations at higher contrast to have a more precise answer to this question. Also it was suggested (Liu et al. 2004) that the spatially localized enhancements and deficits found in the GJ 803 disk at distances larger than $15$ AU from the star might be signposts of ongoing planet formation. Our observations can not infer any information at this large distance from the star. 



\acknowledgments     
We thank the ESO staff for their support during the observations. We also acknowledge
I. Baraffe and A. Burrows for providing informations on their
models and W. Brandner and A. Raga for useful discussions.

\clearpage

\begin{figure}[ht]
\centering
\caption{Theoretical absolute magnitudes in the $K$ and $H$ bands vs. planet mass (in Jupiter mass units) according to the COND models of Baraffe et al. (2003). The light lines (green in on-line version) refer to the $H$ band and the dark lines (black in on-line version) to the $K$ band. The full thin line represents an age of $10$ Myr, the dotted thin line an age of $30$ Myr, the dot-dashed bold line an age of $100$ Myr.
\label{fig1}}
\end{figure}

\begin{figure*}[ht]
\centering
\caption{Two deep images obtained through the $H$ band filter (BD +$2^{\circ}$ $1729$ on the left, BD +$1^{\circ}$ $2447$ on the right). Both images have been processed using a high pass spatial frequency filter. Three point-like sources (A, B and C) are visible at the same position, at around $1$$\arcsec$ from the central star in both images and in all other deep $H$ band images. This clearly indicates that these point-like sources are ghost images produced by the $H$ band filter.}
\label{bd_filter}
\end{figure*}

\clearpage

\begin{figure*}[ht]
\centering
\caption{Left: a point source (marked with a circle) is visible at around $30$ AU from the star SAO $252852$ (galactic latitude -$4.6^{\circ}$). The deep image was processed using a high pass spatial frequency filter. A, B and C indicate the ghost images due to the H band filter (see Fig.\ref{bd_filter}). Right: several point sources (marked with circles) are detected at several tens of AU from the star V343 Nor (galactic latitude -$1.9^{\circ}$). In this picture the deep image is not processed with a high pass spatial frequency filter because all point sources appear at distances larger than $2$$\arcsec$ from the central star. The encircled point sources are probably background stars since both stars are at low galactic latitudes. 
}
\label{sao_cand}
\end{figure*}

\clearpage

\begin{figure*}[ht]
\centering
\caption{{\bf (a)} Deep image of V$343$ Nor ($2\farcs{7}$ x $2\farcs{7}$) obtained in the $Ks$ band ($\sim$ $22$ min) and processed with a high 
pass spatial frequency filter. The PSFs of two $5$M$_{J}$ planets ('artificial planets') were obtained by scaling a non-saturated PSF of the central star. Two artificial planets were placed at $0\farcs{7}$ and $1$$\arcsec$ from the central star. {\bf (b)} The same as (a) but using a different filtering method. {\bf (c)} Contrast expressed in $\Delta M$ (at $5$$\sigma$) vs. angular separation derived from the filtered images shown in (a) and (b). The dashed line represents the detection limit of the non-filtered PSF of the star, the dotted and thin full lines represent the detection limit of the PSF of the star filtering out low spatial frequencies by using the two different methods shown in (a) and (b) (see text). The x-axis projection of the black dot represents the minimum angular separation at which a $5$M$_{J}$ planet should be detectable. The $0\farcs{7}$ value corresponds to the visible planet in the (a) and (b) pictures. The grey scale is arbitrary, the contrast in the picture (b) is better than in picture (a). }
\label{mascia}
\end{figure*}

\clearpage

\begin{figure*}[ht]
\centering
\caption{Planet detection limit (or contrast) expressed in $\Delta M$ (at $5$$\sigma$) vs. angular separation for $9$ observed stars. The dashed line refers to the non-filtered images, while the full thin line and dotted line refer to the images processed by a high spatial frequency filter using two different methods (see text). ET represents the total exposure time. The exposure time of the single frame (DIT) is given in Table \ref{tab3}. No particular criterion is used for the order of the targets in this figure as well as in Fig.\ref{det_lim2}, Fig.\ref{det_lim3} and Fig.\ref{det_lim4}.}
\label{det_lim1}
\end{figure*}

\clearpage

\begin{figure*}[ht]
\centering
\caption {As Fig.\ref{det_lim1} for further $9$ stars. The bumps visible in GJ $799$A and BD $-17^{\circ}$$ 6128$ corresponds to the companion of the binary system.}
\label{det_lim2}
\end{figure*}

\clearpage

\begin{figure*}[ht]
\centering
\caption{As Fig.\ref{det_lim1} for further $9$ stars. }
\label{det_lim3}
\end{figure*}

\begin{figure*}[ht]
\centering
\caption{As Fig.\ref{det_lim1} for further $2$ targets. }
\label{det_lim4}
\end{figure*}

\clearpage

\begin{figure}[ht]
\centering
\caption{Cumulative distribution of the upper limit for the projected star/planet separation calculated with our sample of targets. Thin line: $5$M$_{J}$, bold line: $10$M$_{J}$.
This distribution implies that in $50 \%$ of cases (median value) we have not found any $5$M$_{J}$ planets at distances larger than $14$ AU and no $10$M$_{J}$ planets at distances larger than $8.5$ AU.}
\label{prob}
\end{figure}

\clearpage

\begin{figure}[ht]
\centering
\caption{Detection limit vs. angular separation of one target of the survey obtained with integration times of $5$, $10$ and $22$ min. The two horizontal bars placed at $0\farcs{5}$ mark the expected contrast reachable with SDI/NACO and CHEOPS (VLT planet finder - see text).}
\label{sdi_pot}
\end{figure}


\begin{deluxetable}{cccccccccc}
\setlength{\tabcolsep}{0.08in}
\tabletypesize{\tiny}
\rotate
\tablecaption{Poperties of the central star$^{(a)}$}
\tablewidth{0pt}
\tablehead{ 
 \colhead{Targets} & \colhead{$\alpha$ (2000)  } & \colhead{$\delta$ (2000)  } & \colhead{ST} & \colhead{D} & \colhead{V} &\colhead{Group Ass.} & \colhead{Age} & \colhead{Gal. Lat. }& \colhead{Refer.$^{(b)}$} \\
 &  (h m s)& (${\circ}$ ' '')& & \colhead{pc} & \colhead{(mag)}& &  \colhead{(Myr)} &  \colhead{($^{\circ}$)} &  
}
\startdata
 Hip 2729 & 00 34 51.2& -61 54 58 & K5V & 49.5 & 9.56 & Tuc. Ass.& 35 & -55 & (5),(6),(7)  \\
 HD 17925 & 02 52 32.1& -12 46 11 &K1 & 10.4 &  6.0 &local ass. & 50 & -58 & (8),(12)   \\
 GJ 179 & 04 52 05.7& 06 28 35 &M3.5 & 12.1 &  11.96& & 200 & -23 &(2)   \\
 GJ 182 & 04 59 34.8 & 01 47 01 &M0.5 & 26.7  & 10.1 & & 35 & -24 & (2),(10) \\
 Hip 23309 & 05 00 47.1& -57 15 26 &K7V & 26.3 & 10.09 &Beta Pic.& 12 & -37 &(3),(10),(13) \\
 GJ 207.1 & 05 33 44.8 & 01 56 43 &M2.5 & 16.8 & 11.5  & & 50 & -2 &(1),(2)    \\
 AO Men & 06 18 28.2& -72 02 42 &K5V & 38.5 & 10.09 &Beta Pic. & 12 &-28  &(3),(10),(13)  \\
 BD +2$^{\circ}$ 1729 & 07 39 23.0& 02 01 01 &K7 & 9.8 & 9.82 &local ass. & 100 &12  &(2)   \\
 LQ Hya & 09 32 25.6& -11 11 05 &K0V & 18.3 & 7.80 & & 50 & 28 &(1),(8)    \\
 TWA 6A & 10 18 28.8& -31 50 02 &K7 & 77.0 & 12.0 &TW Hya.& 10 & 21 & (4),(5)   \\
 BD 1$^{\circ}$ 2447 & 10 28 55.5& 00 50 28 &M2 & 7.2 & 9.63 &local ass. & 100 & 47 &(1),(2)  \\
 TWA 5B (2$\arcsec$) & 11 31 55.4& -34 36 27 &M8.5 & 50.0 & - &TW Hya. & 10 & 25 &(4),(5)   \\
 TWA 8A (13$\arcsec$) & 11 32 41.5& -26 51 55 &M2 & 21.0 & (R=11.2) &TW Hya.& 10 & 33 &(4),(5)   \\
 TWA 8B & 11 32 41.5& -26 51 55  &M5 & 21.0 & (R=13.85)&TW Hya.& 10 & 33 &(4),(5) \\
 TWA 9A$^{(c)}$ (9$\arcsec$) & 11 48 24.2& -37 28 49 &M1 & 64.0$^{(d)}$ & 11.3 &TW Hya.& 10 & 24 &(4),(5)   \\
 TWA 9B & 11 48 24.2& -37 28 49 &M1 & 64.0$^{(d)}$ & (R=12.98) &TW Hya.& 10 & 24 &(4),(5)   \\
 SAO 252852 & 14 42 28.1& -64 58 43 &K5V & 16.4 & 8.50  & & 100 & -4 &(2),(9)    \\
 V343 Nor & 15 38 57.6& -57 42 27  & K0V & 39.8 & 8.14 &Beta Pic. & 12 & -2 &(3),(10),(13) \\
 V2306 Oph & 16 30 18.1& -12 39 45 &M3.5 & 4.4 & 10.12 &local ass. & 100 & 24 &(2)    \\
 HD155555 AB$^{(e)}$ & 17 17 25.5& -66 57 02 & G5IV & 31.4 & 7.20 &Beta Pic. & -16 &24  &(3),(10),(13) \\
 HD 155555 C & 17 17 25.5 & -66 57 00 &M4.5 & 31.4 & 12.71 &Beta Pic. & 12 & -16 &(3),(10),(13)  \\
 CD -64$^{\circ}$ 1208 & 18 45 36.9& -64 51 48 &M0 & 29.2 & 9.20 &Beta Pic. & 12 & -24 &(3),(10),(13) \\
 PZ Tel & 18 53 05.9& -50 10 50 &K0 & 49.6 & 8.42  &Beta Pic./Tuc. Ass. & 12 &-21  &(3)   \\
 GJ 799A (2$\farcs{8}$) & 20 41 51.1& -32 26 07 &M4.5 & 10.2 & 11.0 &Beta Pic.& 12 & -36 &(3),(10),(13) \\
 GJ 799B & 20 41 51.1& -32 26 09 &M4.5 & 10.2 & 11.0 &Beta Pic.& 12 & -36 &(3),(10),(13)    \\
 GJ 803 (Au Mic) & 20 45 09.3& -31 20 24 &M0 & 9.9 & 8.61 &Beta Pic. & 12 & -37 &(3),(10),(13)  \\
 BD -17$^{\circ}$ 6128 (2$\farcs{2}$)& 20 56 02.7& -17 10 54 &K7& 47.7 & 10.60 &Beta Pic.& 12 & -35 & (3),(10),(13)  \\
 GJ 813 & 20 57 25.4& 22 21 45 & M2 & 13.6 & 12.0  & & 200 &-15  &(2)  \\
 GJ 890 & 23 08 19.5& -15 24 35 &M0 & 21.8 & 10.84 && 100 & -63 &(1),(2),(11)   \\
 HD 221503 & 23 32 49.4& -16 50 44 & K5 & 13.9 & 8.60 & local ass.& $\ge$ 100-200 & -69 &(1),(2),(9)  \\
\enddata
\tablenotetext{(a)}{Legend, from the left to the right: name of the target, right ascension, declination, spectral type, distance of the target, magnitude in V band, name of the association or group to which the target belongs, age, galactic  latitude, references} 
\tablenotetext{(b)}{Ref. related to the distance and age of the central star: (1) Barrado y Navascu\'es priv. comun.; (2) Montes et al. 2001; (3) Zuckerman et al. 2001a; (4) Webb et al. 1999; (5) Song et al. 2003; (6) Zuckerman \& Webb 2000; (7) Zuckerman et al. 2001b; (8) Wichmann et al. (2003); (9) Poveda et al. (1994); (10) Barrado y Navascues et al. (1999); (11) Barrado y Navascues 1998; (12) Favata et al. (1998); (13) Barrado y Navascues 2001}
\tablenotetext{(c)}{Based on their location in IR color-magnitude or HR diagrams, both TWA9A 
and TWA9B might be older than other members of the TWA 
association whose parallaxes have been measured by Hipparcos 
(Barrado y Navascues, Mohanty \& Jayawardhana (2004))}
\tablenotetext{(d)}{A mean value of trigonometric and photometric distance was used (see (5))}
\tablenotetext{(e)}{Spectroscopic binary (MacIntosh et al. 2001)}
\label{tab1}
\end{deluxetable}
\begin{deluxetable}{cccccccc}
\tabletypesize{\small}
\tablecaption{Infrared magnitude for the central star and the possible planetary companion$^{(a)}$}
\tablewidth{0pt}
\tablehead{
 \colhead{Targets} &  \colhead{$m_{s,K}^{(b)}$} & \colhead{$m_{s,H}^{(b)}$} & \colhead{$m_{({5M_{J},K})}$} & \colhead{$m_{({5M_{J},H})}$} &
\colhead{$\Delta(M_{K})$} & \colhead{$\Delta(M_{H})$} &\colhead{($H$-$K$)$_{5M_{J}}$}\\
 &  \colhead{(mag)} & \colhead{ (mag)} & \colhead{(mag)} & \colhead{(mag)} &
\colhead{(mag)} & \colhead{(mag)} & \colhead{(mag)} 
}
\startdata
 Hip 2729 &   5.96   & 6.13  & 18.60  & 18.75 & 12.64 & 12.62  & -0.02  \\
 HD 17925 &  3.50   & 3.60   & 16.28  & 15.99 & 12.78 & 12.39 &  -0.4  \\
 GJ 179 &   7.77   &  8.01  & 20.24 & 18.31 & 12.47 & 10.30 & -2.17   \\
 GJ 182 &   6.49   & 6.66  & 17.25  &17.40  & 10.76 & 10.74 & -0.02   \\
 Hip 23309  & 6.37   & 6.54   & 15.19  & 15.59 & 8.82 & 9.05 & 0.23    \\
 GJ 207.1 & 7.20   &  7.41  & 17.33  &17.04 & 10.13 & 9.63 &  -0.29   \\
 AO Men &  7.80 & 7.97  & 15.98  & 16.38 & 8.18 & 8.41 & 0.23 \\
 BD +2$^{\circ}$ 1729 & 6.61   & 6.76  & 18.11  & 16.98 & 11.50 & 10.22 &  -1.28  \\
 LQ Hya &  5.97 & 6.07   & 17.51  &17.23  & 11.54 & 11.16 & -0.38   \\
 TWA 6A$^{(c)}$ & 7.97  &  8.17  & 17.35  & 17.65 & 9.38 & 9.48 & 0.15  \\
 BD 1$^{\circ}$ 2447  &  5.58  & 5.78 & 17.44 & 16.31 & 11.86 & 10.53 & -1.33  \\
 TWA 5B (2$\arcsec$)  &  11.5  & 11.95 & 16.41  & 16.71 & 4.91  & 4.76  & -0.15  \\
 TWA 8A (13$\arcsec$)  & 7.44    &  7.72 & 14.53  & 14.83 & 7.09 & 7.11 & 0.14    \\
 TWA 8B &  9.01   &  9.36 & 14.53 & 14.83 & 5.47 & 5.52  &  0.00 \\
 TWA 9A (9$\arcsec$) & 7.68  & 7.95 & 16.95  &17.25  &9.27 & 9.30 & 0.1  \\
 TWA 9B &  9.14   & 9.34  & 16.95  &17.25  & 7.81 & 7.91 & 0.1  \\
 SAO 252852 &  5.68  & 5.85 & 19.22  & 18.09& 13.54 & 12.24 & -1.3   \\
 V343 Nor &  6.31   & 6.41  & 16.05  & 16.46 & 9.74 & 10.05 &  0.31  \\
 V2306 Oph &  5.93   & 6.17  &16.30  & 15.17 & 10.37 & 9.00 & -1.37   \\
 HD155555 AB & 5.62  & 5.7 & 15.54  &15.94  & 9.92 & 10.24 &  0.32  \\
 HD 155555 C & 7.25  & 7.51 & 15.54  & 15.94 & 8.29 & 8.43 & 0.14  \\
 CD -64$^{\circ}$ 1208  &   5.73  & 5.88  & 15.38 &15.78  & 9.65 & 9.90 & 0.25   \\
 PZ Tel &   6.69    & 6.79  & 16.54  &16.94  & 9.85 & 10.15 & 0.3   \\
 GJ 799A (2$\farcs{8}$) &  5.54   & 5.8  & 13.10  & 13.50 & 7.56 & 7.70 & 0.14   \\
 GJ 799B & 5.54  &  5.8 & 13.10  &13.50 & 7.56 & 7.70 & 0.14   \\
 GJ 803 (Au Mic) & 5.00  & 5.15  & 13.03  & 13.44 & 8.03 &  8.29&   0.26 \\
 BD -17$^{\circ}$ 6128 (2$\farcs{2}$) & 7.13   & 7.28 & 16.45  &16.85  & 9.32 &  9.57 & 0.25  \\
 GJ 813 &  7.95  &  8.15  & 20.50 &18.57  & 12.55 &  10.42&  -2.13 \\
 GJ 890 &   7.37    & 7.52  & 19.84 & 18.71 & 12.47 &  11.19& -1.13  \\
 HD 221503 &  5.78  & 5.95   & $\ge$ 20.55  & $\ge$ 18.62 & $\ge$14.77 & $\ge$12.67 & $\ge$ -2.10  \\

\enddata
\tablenotetext{(a)}{Legend, from the left to the right: name of the target,  magnitude of the central star in $K$ band ($m_{s,K}$) and in $H$ band ($m_{s,H}$), magnitude of a potential planet (retrieved from COND models) having a $5$M$_{J}$ mass in $K$ band ($m_{p,K}$) and in $H$ band ($m_{p,H}$), contrast star/planet in $K$ and in $H$ band, color ($H$-$K$).} 
\tablenotetext{(b)}{Calculated from V and spectral type}
\tablenotetext{(c)}{The magnitudes in $H$ and $K$ of all the targets belonging to TWA Hydrae refer to Webb et al. (1999).}
\label{tab2}
\end{deluxetable}

\clearpage

\begin{deluxetable}{cccccc}
\tabletypesize{\scriptsize}
\tablecaption{Observational and instrumental parameters$^{(a)}$}
\tablewidth{0pt}
\tablehead{
\colhead{Targets}   & \colhead{Narrow-band} & \colhead{Broad-band} &  \colhead{DIT. - Broad-band} & \colhead{Exp.Time - Broad-band} & \colhead{Date } \\
 &   &  &  \colhead{(sec)}  & \colhead{(min)} & \colhead{(dd/mm/yyyy)}
}
\startdata
Hip 2729&   NB 2.12& $K_{s}$ & 1.30& $\sim$ 28 & 23/07/2003 \\
HD 17925 & IB 2.09& IB 2.24& 1.70 & $\sim$ 23 & 22/07/2003\\
GJ 179 &  NB 1.64& $H$ & 1.74& $\sim$ 36& 26/12/2003\\
GJ 182 &  NB 2.12& $K_{s}$ & 1.30& $\sim$ 35& 26/12/2003 \\
Hip 23309 &  NB 2.12& $K_{s}$ & 2.06& $\sim$ 20& 23/07/2003 \\
GJ 207.1 &  $K_{s}$/ND Short& $K_{s}$ & 9.00& $\sim$ 30& 08/01/2004 \\
AO Men &  NB 2.12& $K_{s}$ & 2.50 & $\sim$ 20& 17/02/2003 \\
BD +2$^{\circ}$ 1729 &  NB 1.64& $H$ & 1.00& $\sim$ 17& 19/02/2003 \\
LQ Hya &  NB 2.12& $K_{s}$ & 0.80& $\sim$ 23& 09/01/2004 \\
TWA 6A &   NB 2.12& $K_{s}$ & 5.00& $\sim$ 14& 09/01/2004  \\
BD 1$^{\circ}$ 2447 &  $H$/ND Short& $H$ & 0.36& $\sim$ 21& 09/01/2004 \\
TWA 5B (2$\arcsec$)&   NB 2.12& $K_{s}$ & 5.00& $\sim$ 31& 17/02/2003 \\
TWA 8A (13$\arcsec$)&  NB 2.12& $K_{s}$ & 4.20& $\sim$ 19& 17/02/2003 \\
TWA 8B &  $K_{s}$/ND Short& $K_{s}$ & 19.00& $\sim$ 22& 17/02/2003 \\
TWA 9A (9$\arcsec$)&  $K_{s}$/ND Short& $K_{s}$ & 21.00& $\sim$ 25& 18/02/2003 \\
TWA 9B&  $K_{s}$/ND Short& $K_{s}$ & 21.00& $\sim$ 25& 18/02/2003  \\
SAO 252852&  NB 1.64& $H$ & 0.40& $\sim$ 24&21/07/2003  \\
V343 Nor&  NB 2.12& $K_{s}$& 1.90& $\sim$ 22& 19/02/2003 \\
V2306 Oph&  NB 2.12& $K_{s}$ & 1.30& $\sim$ 25& 19/02/2003  \\
HD155555 AB&  NB 2.12& $K_{s}$ & 1.20 & $\sim$ 8& 18/02/2003 \\
HD 155555 C&  NB 2.12& $K_{s}$ & 5.40& $\sim$ 38& 20/07/2003 \\
CD -64$^{\circ}$ 1208&  NB 2.12&  $K_{s}$ & 1.10& $\sim$ 24& 22/07/2003 \\
PZ Tel&   NB 2.12& $K_{s}$ & 1.56& $\sim$ 30& 23/07/2003 \\
GJ 799A (2$\farcs{8}$)&  NB 2.12& $K_{s}$ & 0.58& $\sim$ 23& 21/07/2003 \\
GJ 799B&  NB 2.12& $K_{s}$ & 0.58& $\sim$ 23& 21/07/2003  \\
GJ 803 (Au Mic)&  Ks/ND Short& $K_{s}$ & 0.36& $\sim$ 30& 21/07/2003 \\
BD -17$^{\circ}$ 6128 (2$\farcs{2}$)&  NB 2.12& $K_{s}$ & 2.50& $\sim$ 27& 21/07/2003 \\
GJ 813&   NB 1.64& $H$ & 0.70& $\sim$ 28& 23/07/2003 \\
GJ 890&  NB 2.12& $K_{s}$ & 3.00& $\sim$ 24& 21-22/07/2003 \\
HD 221503&  NB 1.64& $H$ & 0.70& $\sim$ 22& 22/07/2003  \\
\enddata
\tablenotetext{(a)}{Legend, from the left to the right: name of the target, narrow-band filter, broadband-band filter, exposure time of the single frame, total exposure time, observations date.}
\label{tab3}
\end{deluxetable}

\begin{deluxetable}{ccccc}
\tabletypesize{\scriptsize}
\tablecaption{Upper limit of the star/planet projected separation in the cases of 
a $5$$M_{J}$ and a $10$$M_{J}$ planet.}
\tablewidth{0pt}
\tablehead{
  \colhead{Targets}&  \colhead{Age}& \colhead{D}& \colhead{proj. sep.-AU} & \colhead{proj. sep.-AU}\\ 
& (Myr) & (pc)&  \colhead{($5$$M_{J}$)} &  \colhead{($10$$M_{J}$)} }
\startdata
Hip 2729 & 35& 49.5& 64 & 36 \\
GJ 179 & 200& 12.1& 9 & 5\\
GJ 182 & 35& 26.7&19 & 11 \\
Hip 23309 & 12& 26.3&11 & 7  \\
AO Men & 12& 38.5&13  & 9 \\
BD +2$^{\circ}$ 1729 & 100& 9.8& 8& 4 \\
LQ Hya & 50& 18.3&22 & 8 \\
TWA 6A & 10& 77.0&52 & 25 \\
BD 1$^{\circ}$ 2447 & 100& 7.2& 6 & 3 \\
TWA 5B & 10& 50.0& 10 & 7\\
TWA 8A & 10 & 21.0&7 & 3  \\
TWA 8B & 10& 21.0&7 & 2 \\
TWA 9A & 10& 64.0&38 & 19 \\
TWA 9B & 10& 64.0&22 & 13 \\
SAO 252852 & 100& 16.4&19 & 10 \\
V343 Nor & 12& 39.8&24 & 12 \\
V2306 Oph & 100& 4.4&4 & 1 \\
HD 155555 AB & 12&31.4&22 & 11\\
HD 155555 C & 12 & 31.4&13  & 8  \\
CD -64$^{\circ}$ 1208 &12 & 29.2&15  & 9\\
PZ Tel &  12& 49.6&26  & 15 \\
GJ 799 A & 12& 10.2&4  & 2   \\
GJ 799 B & 12& 10.2& 6  & 2  \\
GJ 803 (Au Mic) &  12& 9.9&4  & 2 \\
BD -17$^{\circ}$ 6128 & 12  & 47.7&25 & 14\\
GJ 813 & 200 & 13.6&8  & 4 \\
GJ 890 & 100& 21.8 &26 & 14 \\
HD 221503 & $\ge$ 100-200 & 13.9& 17  & 9 \\

\enddata
\label{tab4}
\end{deluxetable}

\begin{deluxetable}{cc | cc}
\tabletypesize{\scriptsize}
\tablecaption{Statistical Results$^{(a)}$}
\tablewidth{0pt}
\tablehead{
  \colhead{Cases} &  \colhead{AU} & \colhead{Cases} &  \colhead{AU} \\ 
   \colhead{$5$M$_{J}$} &  & \colhead{$10$M$_{J}$}  & }
\startdata
10 cases & 4  $\le$ $r$ $<$ 10  &9 cases & 1  $\le$ $r$ $<$ 5\\
8 cases & 10  $\le$ $r$ $<$ 20  & 8 cases & 5 $\le$ $r$ $<$ 10\\
8 cases & 20  $\le$ $r$ $<$ 40  & 9 cases & 10  $\le$ $r$ $<$  20  \\
2 cases & 40  $\le$ $r$ $<$ 65  & 2 cases & 20 $\le$ $r$ $<$ 36  \\
\enddata
\tablenotetext{(a)}{In the first and third columns the number of cases in which a $5$$M_{J}$ and a $10$$M_{J}$ exo-planet would have been detected in our survey.}
\label{tab5}
\end{deluxetable}


\begin{thebibliography}{}

\bibitem[alonso2004]{alonso2004}
Alonso, R., Brown, T., Torres, G., Latham, D.W., Sozzetti, A., Mandushev, G., Belmonte, J.A., Charbonneau, D., Deeg, H.J., Dunham, E.W., O'Donovan, F.T. \& Stefanik, R.P. 2004,
ApJ, 613, L153

\bibitem[2004]{benz2004}
Alibert, Y., Mordasini, C., Benz, W. 2004,
 A\&A, 417, L25 

\bibitem[baraffe2003]{Baraffe2003}
Baraffe, I., Chabrier, G., Barman, T.S., Allard, F., Hauschild, P.H. 2003,
  A\&A, 402, 701 

\bibitem[barrado1998]{Barrado1998}
Barrado y Navascues, D. 1998,
  A\&A, 339, 831

\bibitem[barrado1999]{Barrado1999bis}
Barrado y Navascues, D., Stauffer, J.R., Song, I., Caillault, J.P. 1999,
 ApJ, 520, L123

\bibitem[barrado1999]{Barrado1999}
Barrado y Navascues, D. 1999,
  A\&A, 402, 701 

\bibitem[barrado2001]{Barrado2001}
Barrado y Navascues, D. 2001,
  ASP, Conf. Vol. 244, , Ed. Jayawardhana R. and Greene T, San Francisco

\bibitem[barrado2004]{Barrado2004}
Barrado y Navascues, Mohanty \& Jayawardhana 2004,
in prep.

\bibitem[black1980]{black1980}
Black, D.C. 1980,
Icarus, 43, 293

\bibitem[bouchy2004]{bouchy2004}
Bouchy, F., Pont, F., Santos, N.C., Melo, C., Mayor, M., Queloz, D., Udry, S. 2004, A\&A, 421, L13

\bibitem[brandner2000]{brandner2000}
Brandner, W., Zinneker, H., Alacal\'a, J.M., Allard, F., Covino, E., Frink, S., K\"ohler, R., Kunkel, M., Monetti, A, Schweitzer, A. 2000,
AJ, 120, 950


\bibitem[1997]{Burrows1997} 
Burrows, A., Marley, M., Hubbard, W.B., Lunine, J.I., Guillot, T., Saumon, D., Freedman, R.,
Sudarsky, D., Sharp, C. 1997
  ApJ, 491, 856

\bibitem[Charbon2000]{charbon2000}
Charbonneau, D., Brown, T.M., Latham, D.W., Mayor, M. 2000
ApJ, 529, L45, 2000

\bibitem[Close al.(2004)]{Close2004}
Close, L.M., Lenzen, R., Biller, B., Brandner, W., Hartung M., 2004,
  ESO Workshop ``Science with AO'', to be published

\bibitem[chauvin2004]{chauvin2004}
Chauvin, G., Lagrange, A.-M., Dumas, C., Zuckerman, B., Mouillet, D., Song, I., Beuzit, J.-L. and Lowrance, P. 2004,
A\&AL, 425, L29  


\bibitem[eggen1996]{eggen1996}
Eggen, O.J. 1996
 AJ, 111, 466

\bibitem[favata1998]{favata1998}
Favata, F., Micela, G., Sciortino, S. \& D'Antona, F. 1998
 A\&A, 335, 218

\bibitem[ford2001]{ford2001} 
Ford, E.B., Havlickova, M., Rasio, F.A. 2001
Icarus, 150, 303


\bibitem[gratton2004]{gratton2004}
Gratton, R., Feldt, M., Schmid, H.M., Brandner, W., Hippler, S., Neuh\"auser, R., Quirrenbach, A., Desidera, S., Turatto, M., Stam, D.M. 2004,
SPIE, 5492, 1010

\bibitem[haisch2001]{haisch2001}
Haisch, K.E., Lada, E.A. \& Lada, C.J. 2001,
ApJ, 553, L153

\bibitem[kaisler2003]{kaisler2003}
Kaisler, D., Zuckerman, B., Becklin, E. 2003,
ASP Conf. Series, 294, 91


\bibitem[kalas2004]{kalas2004}
Kalas, P., Liu, M.C., Matthews, B.C. 2004, Science, 303, 1990


\bibitem[konacki2004]{konacki2004}
Konacki, M., Torres, G., Sasselov, D., Pietrzynski, G., Udalski, A., Jha, S., Ruiz, M.T., Gieren, W., Minniti, D., 2004
ApJL, submitted (astro-ph/0404541)

\bibitem[jeffries1995]{jeffries1995}
Jeffries, R.D. 1995,
 MNRAS, 273, 559

\bibitem[1998]{Lenzen1998}
Lenzen R. et al., 1998, 
  {\it SPIE}, 3354, 606

\bibitem[Lenzen al.(2004)]{Lenzen2004}
Lenzen, R., Close, L.M., Brandner, W., Hartung M., Biller, B.,2004,
  ESO Workshop ``Science with AO'', to be published

\bibitem[1993]{Lissauer1993}
Lissauer, J.J. 1993,
\araa,31, 129


\bibitem[liu2004]{liu2004}
Liu, M.C. 2004, Science, 305, 1442

\bibitem[2002]{Luhman2002}
Luhman, K.L. \& Jayawardhana, R. 2002,
  ApJ, 566, 1132

\bibitem[2001]{Macintosh2001}
Macintosh, B.A., Max, C., Zuckerman, B., Becklin, E.E., Kaisler, D., Lowrance, P., Weinberger, A. 2001, 
  ASP Conferences Series, 244, 309




\bibitem[McCarty2004]{mccarty2004}
McCarty, C. \& Zuckerman, B. 2004,
ApJ, 127, 2871

\bibitem[Marcy2003]{marcy2003} 
Marcy, G.W., Butler, R.P., Fischer, D.A., Vogt, S.S. 2003
ASP Conf., 294, 1

\bibitem[1995]{Mayor1995}
Mayor, M. \& Queloz, D. 1995,
  ApJ, 378, 355


\bibitem[2001]{Montes2001}
Montes, D., L\'opez-Santiago J., G\'alvez, M.C., Fern\'andez-Figueroa, M.J., De Castro E., Cornide, M. 2001,
  MNRAS, 328, 45


\bibitem[neuhauser2002]{neuhauser2002}
Neuh\"auser, R., Brandner, W., Alves, J., Joergens, V., Comeron, F. 2002,
A\&A, 384, 999

\bibitem[Neuhauser1997]{neuh1997}
Neuh\"auser, R., Brandner, W., Eckart, A., Guenther, E.W., Alves, J.,  Ott, Th., Hu\'elamo, N., Fern\'andez, M. 2003,
A\&A, 354, L9

\bibitem[Neuhauser2003]{neuh2003}
Neuh\"auser, R., Guenther, E.W., Alves, J., Hu\'elamo, N., Ott, Th., Eckart, A. 2003,
Astron. Nachr., 324, 535


\bibitem{Papa2001} 
Papaloizou J.C.B. \& Terquem, C. 2001,
MNRAS, 325, 221

\bibitem[perryman1997]{perryman1997}
Perryman, M.A.C., Lindegren, L., Kovalesky, J. et al. 1997,
A\&A, 323, L49 

\bibitem[Pollack1996]{Pollack1996}
Pollack, J.B., Hubickyj, O., Bodenheimer, P., Lissauer, J.L., Podolak, M., Greenzweig, Y. 1996
Icarus, 124, 62

\bibitem[Pont2004]{Pont2004}
Pont, F., Bouchy, F., Queloz, D., et al. 2004
A\&A, in press

\bibitem[poveda1994]{poveda1994}
Poveda, A., Herrera, M.A., Allen, C., Cordero., G., Lavalley, C. 1994,
  RMxAA, 28, 43

\bibitem[Racine et al.(1999)]{Racine99}
Racine R., Walker G.A.H., Nadeau D., Doyon R., Marois C. 1999,
  \pasp, 111, 587

\bibitem[rasio1996]{Rasio1996} 
Rasio, F.A., Ford, E.B. 1996,
 Science, 274, 954


\bibitem[Rosenthal1996]{Rosenthal1996} 
Rosenthal, E.D., Gurwell, M.A., Ho, P.T.P 1996,
Nature, 384, 243
 
\bibitem[2000]{Rousset2000}
Rousset G. et al. 2000, 
Proc. SPIE, 4007, 72

\bibitem[song2003]{song2003}
Song, I., Zuckerman, B., Bessel, M.S. 2003,
ApJ, 599, 342

\bibitem[1995]{stauffer1995}
Stauffer, J.R., Hartmann, L.W., Barrado y Navascues, D. 1995, 
 ApJ, 454, 910

\bibitem[thommes1999]{thommes1999}
Thommes E.W., Duncan, M.J., Levison, H. 1999,
Nature, 402, 635 

\bibitem[thommes2002]{thommes2002}
Thommes E.W., Duncan, M.J., Levison, H. 2002,
AJ, 123, 2862

\bibitem[torres2004]{torres2004}
Torres, G., Konacki, M., Sasselov, D.D., Jha, S. 2004,
The search for other worlds: Fourteenth Astrophysics Conference - AIP Conference Proceedings, 713, 165

\bibitem[veras2004]{veras2004}
Veras, D. \& Armitage, P.J. 2004,
MNRAS, 347, 613

\bibitem[webb1999]{webb1999}
Webb, R.A., Zuckerman, B., Platais, I, Pacience, J., WHite, R.J., Schwartz, M.J., McCarthy, C. 1999,
ApJ, 559, 388

\bibitem[1996]{Weiden1996}
Weidenschilling, S. J. \& Marzari, F. 1996,
  Nature, 384, 619

\bibitem[2003]{Wichmann2003}
Wichmann, R., Schmitt, J.H.M.M., Hubrig, S. 2003,
  A\&A, 399, 983

\bibitem[zuke2000]{zuke2000}
Zuckerman, B. \& Webb, R.A. 2000,
ApJ, 535, 959

\bibitem[zuke2001]{zuke2001}
Zuckerman, B., Song, I., Bessel, M.S., Webb, R.A. 2001a,
ApJ, 562, L87

\bibitem[zuke2001b]{zuke2001b}
Zuckerman, B., Song, I., Webb, R.A. 2001b,
ApJ, 559, 388


\end{thebibliography}
\end{document}